\UseRawInputEncoding
\documentclass[epj]{svjour}
\usepackage{graphics}

\usepackage{amssymb}
\usepackage{amsmath}
\usepackage{bm}
\usepackage{mathptmx}

\begin{document}
\title{One underlying mechanism for two piezoelectric effects in the octonion spaces}
\author{Zi-Hua Weng\inst{1} \inst{2}
\thanks{\emph{Email address: } xmuwzh@xmu.edu.cn }%
}                     
%
%
\institute{School of Aerospace Engineering, Xiamen University, Xiamen 361005, China
\and College of Physical Science and Technology, Xiamen University, Xiamen 361005, China}
\date{Received: date / Revised version: date}
%
\abstract{
The paper aims to apply the algebra of octonions to explore the contributions of external derivative of electric moments and so forth on the induced electric currents, revealing a few major influencing factors relevant to the direct and inverse piezoelectric effects. J. C. Maxwell was the first to adopt the algebra of quaternions to describe the physical quantities of electromagnetic fields. The contemporary scholars utilize the quaternions and octonions to research the physical properties of electromagnetic and gravitational fields. The application of octonions is able to study the physical quantities of electromagnetic and gravitational fields, including the octonion field strength, field source, linear moment, angular moment, torque and force. When the octonion force is equal to zero, it is capable of achieving eight independent equations, including the force equilibrium equation, fluid continuity equation, current continuity equation, and second-precession equilibrium equation and others. One of inferences derived from the second-precession equilibrium equation is that the electric current and derivative of electric moments are able to excite each other. The external derivative of electric moments can induce the electric currents. Meanwhile the external electric currents are capable of inducing the derivative of electric moments. The research states that this inference can be considered as the underlying mechanism for the direct and inverse piezoelectric effects. Further the second-precession equilibrium equation is able to predict several new influencing factors of direct and inverse piezoelectric effects.
} 
\maketitle

\section{\label{sec:level1}Introduction}

What is the underlying mechanism of direct and inverse piezoelectric effects? Are there a few new influencing factors for the direct and inverse piezoelectric effects? Can the spatial dimension impact the direct and inverse piezoelectric effects? For many years, these problems have plagued scholars and engineers. They introduced a variety of detection methods, exploring deeply the possible theoretical basis behind the direct and inverse piezoelectric effects, attempting to expand the scope of application of the piezoelectric effects. It was not until the emergence of field theory, described with the algebra of octonions, that these puzzles were answered to a certain extent. The field theory reveals that the electric current and derivative of electric moments are interrelated rather than irrelevant. The external derivative of electric moments is able to induce the generation of electric current, while the external electric current is capable of inducing the derivative of electric moments. This inference can be considered as the underlying mechanism of direct and inverse piezoelectric effects.

In 1880, J. Curie and P. Curie discovered the tourmalines possess the direct piezoelectric effects \cite{curie,lu}. And they verified the inverse piezoelectric effects through experiments in 1881, attaining the direct and inverse piezoelectric constants \cite{kochin}. More than half a century later, the direct and inverse piezoelectric effects have been found in many practical applications. The barium-titanate pickups were born in the 1940s, while the lead zirconate titanates were successfully developed in the 1950s. Since then, the piezoelectric ceramics have entered a new stage of development \cite{mazumdar}. The piezoelectric ceramics experience continuous improvements, the binary, ternary and quaternary piezoelectric ceramics came into being \cite{wieland}. These materials have excellent properties and wide application.

The piezoelectric effects \cite{lei} play an important role in the sound generation and detection, high-voltage generation, electrical frequency generation, microbalance, and ultra-fine focusing of optical devices and others. The appearance of ternary and quaternary piezoelectric ceramics \cite{rampal} has greatly increased the application range of piezoelectric ceramics, including the ceramic filter, ceramic frequency discriminator, electro-acoustic transducer, medical ultrasonic probe, underwater acoustic transducers, surface acoustic wave devices, electro-optical device, infrared detection device, smart material, piezoelectric gyroscope and so forth.

Many materials have also been found to be piezoelectric, including the wood, ramie, bamboo wool, animal bones, skin, blood vessels and so forth. Subsequently, it was found that the synthetic polymers \cite{aziz} possess the piezoelectric properties, while the electro-polarized polyvinylidene fluoride \cite{marshall,halabi} has strong piezoelectric properties. In recent years, many polymer materials with piezoelectric effects have been developed by the synthetic methods, and these polymer materials are one of important types of artificial muscles.

However, the theoretical research, relevant to the underlying mechanism of direct and inverse piezoelectric effects, has made slow progress. In 1910, W. Voigt published his inference that only 20-odd classes of natural crystal without symmetry centers may possess piezoelectric effects \cite{voigt}. And the piezoelectric coefficient of natural crystal can be expressed rigorously by 18 possible components of tensor. At present, the scholars continue to make in-depth theoretical research on the direct and inverse piezoelectric effects.

Through the preceding analysis and comparison, it can be seen that there are some deficiencies relevant to the study of piezoelectric effects, in the classical electromagnetic theory. This restricts the further understanding and application of piezoelectric effects.

1) Interference parameters. The scholars utilize the electric polarization, crystal point group, temperature, frequency and others to analyze the physical characteristics of piezoelectric effects. However, these existing research methods cannot study the piezoelectric properties, from the perspective of several major parameters of piezoelectric materials. At present, there are so many influencing factors to exert an impact on the piezoelectric coefficients. Some secondary factors have interfered with the studies of piezoelectric characteristics.

2) Derivative of electric moments. In the past, the derivative of electric moments was considered as an isolated physical quantity. It is merely related to the electric moment and time, but not to other physical quantities. In particular, the derivative of electric moments is assumed to be independent of the electric currents. This point of view is not conducive to the exploration of piezoelectric effects from the perspective of internal factors. Either it is not beneficial to essentially expanding the application range of piezoelectric effects.

3) Spatial dimension independence. The existing field theories are unable to analyze the influence of spatial dimensions on the piezoelectric effects. In other words, the existing theoretical analysis claims that there are piezoelectric effects of various spatial dimensions. The piezoelectric effects are not limited by any spatial dimension. The spatial dimension is not a factor affecting the piezoelectric effects. As a result, this point of view cannot explain why the piezo-proteins \cite{lwang,myma} must be folded into two- or three-dimensional shapes.

Apparently, the classical field theory cannot effectively study the relationship between the derivative of electric moments and electric current. Either it is incapable of revealing the contribution of magnetic moments and spatial dimensions and others on the derivative of electric moments and electric currents. The disadvantage restricts the further understanding and application of the derivative of electric moments and electric currents. The limitation confines the scope of application of the piezoelectric effects in the classical field theory.

In sharp contrast with the above analysis, the electromagnetic and gravitational theories described with the octonions \cite{xq1} (the octonion field theory for short, temporarily) are able to solve some puzzles originated from the classical field theory, attempting to improve the theoretical explanation related to the underlying mechanism of piezoelectric effects.

J. C. Maxwell first utilized the algebra of quaternions to describe the physical properties of electromagnetic fields. This method encourages the subsequent scholars to apply the quaternions and octonions \cite{xq2,xq3} to explore the electromagnetic fields \cite{xq4}, gravitational fields \cite{kansu1,kansu2}, wave functions \cite{xq5}, black holes, dark matter \cite{xq7}, strong nuclear fields \cite{xq8}, weak nuclear fields \cite{xq9}, Gauge field \cite{xq10}, Hall effect,
dyons \cite{kansu3,kansu4}, hydromechanics \cite{xq12}, quantum computing \cite{xq13}, and plasmas \cite{kansu5} and others.

The application of octonions is able to describe simultaneously the electromagnetic and gravitational fields. When the octonion force (in Section 2) is equal to zero, it is capable of achieving eight equations independent of each other. One of eight equations reveals some influencing factors, such as the derivative of electric moments and electric current, are interrelated. The derivative of electric moments can directly exert a significant impact on the electric currents within piezoelectric materials, vice versa.

In the octonion field theory, the research relevant to the piezoelectric effects possesses a few important characteristics as follows.

1) Main parameters. When the contribution of field strength can be neglected, the curl of magnetic moments, divergence of magnetic moments and second-torque are main influencing factors of piezoelectric effects. It is found that there are a few correlations among the curl of magnetic moments, divergence of magnetic moments and second-torque. The divergence of magnetic moments and second-torque can respectively exert an impact on the curl of magnetic moments. Under certain conditions, the curl of magnetic moments is related to the curl of electric currents, while the second-torque is related to the curl of electric moments. The relationship between the curl of magnetic moments and second-torque can be simplified into that between the electric current and derivative of electric moments. The appearance of derivative of electric moments can induce the generation of electric currents, vice versa.

2) Induced electric currents. The action of external forces will alter the shapes of piezoelectric materials. In other words, the external force will transform the electric moments of piezoelectric materials, varying the derivative of electric moments. The existence of derivative of electric moments will induce a weak electric current within piezoelectric materials. Some different electric charges may appear on the surfaces of piezoelectric materials, in case the external electric circuit is disconnected. On the contrary, the electric moments of piezoelectric materials will be transformed, under the action of external electric currents, altering the derivative of electric moments. The variation of derivative of electric moments may lead to the change of area or volume of piezoelectric materials.

3) Spatial dimension dependence. The theoretical analysis states that the spatial dimension has a direct impact on the piezoelectric effects. For the piezoelectric materials, there are two- and three-dimensional piezoelectric effects. However, in case the spatial dimension is equal to 1, the term related to the electric currents will disappear, in the equation relevant to the electric current and derivative of electric moments. That is, the one-dimensional piezoelectric effect is the lowest for all of integer spatial dimensions, and even there is no one-dimensional piezoelectric effect in the strict sense. The spatial dimension is one quite important factor affecting the piezoelectric effects. The point of view claims that the piezo-proteins \cite{xzfang,bagriantsev} tend to be folded into two- or three-dimensional shapes, enhancing relevant piezoelectric coefficients.

By means of the octonion field theory, this paper explores the causes and related influencing factors of piezoelectric effects, from the perspective of the relationship between the derivative of electric moments and electric current. The derivative of electric moments and electric current are not isolated physical quantities, from second-procession equilibrium equation. In particular, the external derivative of electric moments is able to induce the generation of electric currents. Meanwhile the external electric currents are capable of inducing the derivative of electric moments.

From the algebra of octonions, it is able to describe the physical quantities of gravitational and electromagnetic fields, including the octonion angular momentum, torque, and force and so forth \cite{weng1}. When the octonion force is equal to zero, it can be separated into eight equations independent of each other. Four of them are the equilibrium equations, the rest are the continuity equations. The eight equilibrium or continuity equations are essentially consistent. The three of eight equations had been validated by the experiments, those are, the force equilibrium equation, fluid continuity equation, and current continuity equation. As a result, the other five equations must also be correct in theory. It is found that the second-force equilibrium equation \cite{weng2} has been proved by the experiments. A few inferences derived from the second-force equilibrium equation are consistent with the experimental results \cite{deng,ober}. After finding the verification experiments of the second-force equilibrium equation, we have more confidence in the second-precession equilibrium equation.

\section{Octonion angular momentum}

The octonion space, $\mathbb{O}$ , can be separated into several subspaces independent of each other, including $\mathbb{H}_g$ and $\mathbb{H}_{em}$. The subspace $\mathbb{H}_g$ is one quaternion space, which is fit for describing the physical properties of gravitational fields. And the subspace $\mathbb{H}_{em}$ is appropriate for depicting the physical properties of electromagnetic fields (Table 1). Under some appropriate conditions, the subspace $\mathbb{H}_{em}$ can be transformed into another quaternion space $\mathbb{H}_e$ . The quaternion space $\mathbb{H}_e$ is independent of the quaternion space $\mathbb{H}_g$ . It implies that the octonion space, $\mathbb{O}$ , is able to explain the physical properties of gravitational and electromagnetic fields simultaneously.

In the octonion space, the octonion angular momentum $\mathbb{L}$ is written as (see Ref.[32]),
\begin{eqnarray}
\mathbb{L} = ( \mathbb{R} + k_{rx} \mathbb{X} )^\times \circ \mathbb{P} ~.
\end{eqnarray}

In the above equation, $\mathbb{R}$ is the octonion radius vector, $\mathbb{X}$ is the octonion integrating function of field potential, and $\mathbb{P}$ is the octonion linear momentum. $\mathbb{R} = \mathbb{R}_g + k_{eg} \mathbb{R}_e$ . $\mathbb{R}_g = i r_0 \textbf{d}_0 + \Sigma r_k \textbf{d}_k$ , is the component of radius vector $\mathbb{R}$ in the quaternion space $\mathbb{H}_g$. $\mathbb{R}_e = i R_0 \textbf{D}_0 + \Sigma R_k \textbf{D}_k$ , is the component of radius vector $\mathbb{R}$ in the subspace $\mathbb{H}_{em}$ . $\mathbb{X} = \mathbb{X}_g + k_{eg} \mathbb{X}_e$ . $\mathbb{X}_g = i x_0 \textbf{d}_0 + \Sigma x_k \textbf{d}_k$ , is the component of $\mathbb{X}$ in the quaternion space $\mathbb{H}_g$. $\mathbb{X}_e = i X_0 \textbf{D}_0 + \Sigma X_k \textbf{D}_k$ , is the component of $\mathbb{X}$ in the subspace $\mathbb{H}_{em}$. $\mathbb{P} = \mathbb{P}_g + k_{eg} \mathbb{P}_e$. $\mathbb{P}_g = i p_0 \textbf{d}_0 + \Sigma p_k \textbf{d}_k$ , is the component of linear momentum $\mathbb{P}$ in the quaternion space $\mathbb{H}_g$ . $\mathbb{P}_e = i P_0 \textbf{D}_0 + \Sigma P_k \textbf{D}_k$ , is the component of linear momentum $\mathbb{P}$ in the subspace $\mathbb{H}_{em}$. Herein $\times$ is the complex conjugate. $k_{eg}$ and $k_{rx}$ are two coefficients, meeting the demand of the dimensional homogeneity. $i$ is the imaginary unit. $r_j$ , $R_j$ , $x_j$, $X_j$ , $p_j$, and $P_j$ are all real. $r_0 = v_0 t$. $v_0$ is the speed of light. $t$ is the time. $\textbf{d}_j$ is the basis vector in the quaternion space $\mathbb{H}_g$ , while $\textbf{D}_j$ is the basis vector in the subspace $\mathbb{H}_{em}$ . $\textbf{D}_j = \textbf{d}_j \circ \textbf{D}_0$. $\textbf{d}_0 = 1$. $\textbf{d}_k^2 = -1$. $\textbf{D}_j^2 = -1$. $\circ$ denotes the octonion multiplication. $j = 0, 1, 2, 3$. $k =1, 2, 3$.

Further, the octonion angular momentum $\mathbb{L}$ can be expanded to
\begin{eqnarray}
\mathbb{L} = \mathbb{L}_g + k_{eg} \mathbb{L}_e  ~ ,
\end{eqnarray}
where $\mathbb{L}_g = L_{10} + i \textbf{L}_1^\textrm{i} + \textbf{L}_1$ , is the component of $\mathbb{L}$ in the quaternion space $\mathbb{H}_g$ . $\mathbb{L}_e = \textbf{L}_{20} + i \textbf{L}_2^\textrm{i} + \textbf{L}_2$ , is the component of $\mathbb{L}$ in the subspace $\mathbb{H}_{em}$ . $\textbf{L}_1$ is the angular momentum. And $\textbf{L}_1^\textrm{i}$ is called as the mass moment temporarily. $\textbf{L}_2$ is the magnetic moment, while $\textbf{L}_2^\textrm{i}$ is the electric moment. $\textbf{L}_1 = \Sigma L_{1k} \textbf{d}_k$ , $\textbf{L}_1^\textrm{i} = \Sigma L_{1k}^\textrm{i} \textbf{d}_k$ . $\textbf{L}_{20} = L_{20} \textbf{D}_0$, $\textbf{L}_2 = \Sigma L_{2k} \textbf{D}_k$ , $\textbf{L}_2^\textrm{i} = \Sigma L_{2k}^\textrm{i} \textbf{D}_k$ . $L_{1j}$ , $L_{1k}^\textrm{i}$ , $L_{2j}$, and $L_{2k}^\textrm{i}$ are all real.

In the octonion space, the octonion torque $\mathbb{W}$ will be written as,
\begin{eqnarray}
\mathbb{W} = - v_0 ( i \mathbb{F} / v_0 + \lozenge ) \circ \{ ( i \mathbb{V}^\times / v_0) \circ \mathbb{L} \} ~,
\end{eqnarray}
where $\mathbb{F}$ is the octonion field strength. $\mathbb{F} = \mathbb{F}_g + k_{eg} \mathbb{F}_e$ . $\mathbb{F}_g = f_0 + \textbf{f}$ , is the gravitational strength in the quaternion space $\mathbb{H}_g$ . If we choose the equation, $f_0 = 0$, for the gauge equation of gravitational fields, there is, $\textbf{f} = i \textbf{g} / v_0 + \textbf{b}$. $\textbf{g}$ is the gravitational acceleration, while $\textbf{b}$ is called as the gravitational precessional-angular-velocity temporarily. $\mathbb{F}_e = \textbf{F}_0 + \textbf{F}$ , is the electromagnetic strength in the subspace $\mathbb{H}_{em}$ . In case we choose the equation, $F_0 = 0$, for the gauge equation of electromagnetic fields, there is, $\textbf{F} = i \textbf{E} / v_0 + \textbf{B}$ . $\textbf{E}$ is the electric field intensity, while $\textbf{B}$ is the magnetic induction intensity. $\textbf{f} = \Sigma f_k \textbf{d}_k$. $\textbf{F} = \Sigma F_k \textbf{D}_k$ . $\textbf{F}_0 = F_0 \textbf{D}_0$ . $f_0$ and $F_0$ are all real. $f_k$ and $F_k$ both are complex numbers. $\lozenge$ is the quaternion operator. $\lozenge = \partial_0 + \nabla$ . $\nabla = \Sigma \partial_k \textbf{d}_k$ , $\partial_j = \partial / \partial r_j$.

In the above equation, the octonion torque $\mathbb{W}$ can be expanded to
\begin{eqnarray}
\mathbb{W} = \mathbb{W}_g + k_{eg} \mathbb{W}_e  ~ ,
\end{eqnarray}
where $\mathbb{W}_g = i W_{10}^\textrm{i} + W_{10} + i \textbf{W}_1^\textrm{i} + \textbf{W}_1$ , is the component of $\mathbb{W}$ in the quaternion space $\mathbb{H}_g$ . $\mathbb{W}_e = i \textbf{W}_{20}^\textrm{i} + \textbf{W}_{20} + i \textbf{W}_2^\textrm{i} + \textbf{W}_2$ , is the component of $\mathbb{W}$ in the subspace $\mathbb{H}_{em}$ . The terms of octonion torque $\mathbb{W}$ are seized of their own physical meanings respectively in the Table 1. In particular, $W_{10}^\textrm{i}$ is the energy. $\textbf{W}_1^\textrm{i}$ is the torque, including the gyroscopic torque. $\textbf{W}_{20}^\textrm{i}$ and $\textbf{W}_2^\textrm{i}$ are called as the second-energy and second-torque respectively and temporarily. $\textbf{W}_1 = \Sigma W_{1k} \textbf{d}_k$ , $\textbf{W}_1^\textrm{i} = \Sigma W_{1k}^\textrm{i} \textbf{d}_k$. $\textbf{W}_{20}^\textrm{i} = W_{20}^\textrm{i} \textbf{D}_0$, $\textbf{W}_{20} = W_{20} \textbf{D}_0$ , $\textbf{W}_2 = \Sigma W_{2k} \textbf{D}_k$ , $\textbf{W}_2^\textrm{i} = \Sigma W_{2k}^\textrm{i} \textbf{D}_k$ . $W_{1j}$ , $W_{1j}^\textrm{i}$ , $W_{2j}$ , and $W_{2j}^\textrm{i}$ are all real.

\begin{table}[h]
\centering
\caption{Some physical meanings of eight components of the octonion torque, $\mathbb{W}$ , in the gravitational and electromagnetic fields.}
\begin{tabular}{@{}lll@{}}
\hline\noalign{\smallskip}
term                         &    definition                                                                   &  subspace           \\
\noalign{\smallskip}\hline\noalign{\smallskip}
$W_{10}^\textrm{i}$          &    energy                                                                       &  $\mathbb{H}_g$     \\
$W_{10}$                     &    divergence of angular momentum $\textbf{L}_1$                                &  $\mathbb{H}_g$     \\
$\textbf{W}_1^\textrm{i}$    &    torque                                                                       &  $\mathbb{H}_g$     \\
$\textbf{W}_1$               &    curl of angular momentum $\textbf{L}_1$                                      &  $\mathbb{H}_g$     \\
$\textbf{W}_{20}^\textrm{i}$ &    second-energy                                                                &  $\mathbb{H}_{em}$  \\
$\textbf{W}_{20}$            &    divergence of magnetic moment $\textbf{L}_2$                                 &  $\mathbb{H}_{em}$  \\
$\textbf{W}_2^\textrm{i}$    &    second-torque                                                                &  $\mathbb{H}_{em}$  \\
$\textbf{W}_2$               &    curl of magnetic moment $\textbf{L}_2$                                       &  $\mathbb{H}_{em}$  \\
\noalign{\smallskip}\hline
\end{tabular}
\end{table}

In the octonion space, the octonion force $\mathbb{N}$ can be written as,
\begin{eqnarray}
\mathbb{N} = - ( i \mathbb{F} / v_0 + \lozenge ) \circ \{ ( i \mathbb{V}^\times / v_0) \circ \mathbb{W} \} ~,
\end{eqnarray}
where $\mathbb{N} = \mathbb{N}_g + k_{eg} \mathbb{N}_e$ . $\mathbb{N}_g = i N_{10}^\textrm{i} + N_{10} + i \textbf{N}_1^\textrm{i} + \textbf{N}_1$ , is the component of $\mathbb{N}$ in the quaternion space $\mathbb{H}_g$ . $\mathbb{N}_e = i \textbf{N}_{20}^\textrm{i} + \textbf{N}_{20} + i \textbf{N}_2^\textrm{i} + \textbf{N}_2$, is the component of $\mathbb{N}$ in the subspace $\mathbb{H}_{em}$ . The terms of octonion force $\mathbb{N}$ are in possession of their own physical meanings respectively (Table 2). Especially, $N_{10}$ is the power. $\textbf{N}_1^\textrm{i}$ is the force, including the Magnus force. $\textbf{N}_{20}$ and $\textbf{N}_2^\textrm{i}$ are called as the second-power and second-force respectively and temporarily. $\textbf{N}_1 = \Sigma N_{1k} \textbf{d}_k$ , $\textbf{N}_1^\textrm{i} = \Sigma N_{1k}^\textrm{i} \textbf{d}_k$. $\textbf{N}_{20}^\textrm{i} = N_{20}^\textrm{i} \textbf{D}_0$ , $\textbf{N}_{20} = N_{20} \textbf{D}_0$ . $\textbf{N}_2 = \Sigma N_{2k} \textbf{D}_k$ , $\textbf{N}_2^\textrm{i} = \Sigma N_{2k}^\textrm{i} \textbf{D}_k$ . $N_{1j}$ , $N_{1j}^\textrm{i}$ , $N_{2j}$ , and $N_{2j}^\textrm{i}$ are all real.

\begin{table}[h]
\centering
\caption{Some physical meanings of eight components of the octonion force, $\mathbb{N}$ , in the gravitational and electromagnetic fields.}
\begin{tabular}{@{}lll@{}}
\hline\noalign{\smallskip}
term                         &    definition                                                                        &  subspace           \\
\noalign{\smallskip}\hline\noalign{\smallskip}
$N_{10}^\textrm{i}$          &    divergence of torque $\textbf{W}_1^\textrm{i}$                                    &  $\mathbb{H}_g$     \\
$N_{10}$                     &    power                                                                             &  $\mathbb{H}_g$     \\
$\textbf{N}_1^\textrm{i}$    &    force                                                                             &  $\mathbb{H}_g$     \\
$\textbf{N}_1$               &    derivative of torque $\textbf{W}_1^\textrm{i}$                                    &  $\mathbb{H}_g$     \\
$\textbf{N}_{20}^\textrm{i}$ &    divergence of second-torque $\textbf{W}_2^\textrm{i}$                             &  $\mathbb{H}_{em}$  \\
$\textbf{N}_{20}$            &    second-power                                                                      &  $\mathbb{H}_{em}$  \\
$\textbf{N}_2^\textrm{i}$    &    second-force                                                                      &  $\mathbb{H}_{em}$  \\
$\textbf{N}_2$               &    derivative of second-torque $\textbf{W}_2^\textrm{i}$                             &  $\mathbb{H}_{em}$  \\
\noalign{\smallskip}\hline
\end{tabular}
\end{table}

When the octonion force is equal to zero, that is, $\mathbb{N} = 0$, the above equation can be separated into,
\begin{eqnarray}
&& i N_{10}^\textrm{i} + N_{10} + i \textbf{N}_1^\textrm{i} + \textbf{N}_1 = 0 ~,
\\
&& i \textbf{N}_{20}^\textrm{i} + \textbf{N}_{20} + i \textbf{N}_2^\textrm{i} + \textbf{N}_2 = 0 ~,
\end{eqnarray}
or
\begin{eqnarray}
&& N_{10}^\textrm{i} = 0 ~,~ N_{10} = 0 ~,
\\
&& \textbf{N}_1^\textrm{i} = 0 ~,~ \textbf{N}_1 = 0 ~,
\\
&& \textbf{N}_{20}^\textrm{i} = 0 ~,~ \textbf{N}_{20} = 0 ~,
\\
&& \textbf{N}_2^\textrm{i} = 0 ~,~ \textbf{N}_2 = 0 ~.
\end{eqnarray}

The above analysis states that the eight components perpendicular to each other will all be zero, when one octonion physical quantity is equal to zero. This is similar to the case of the three-dimensional vector space in the classical mechanics. When a three-dimensional vector is equal to zero, each of three components orthogonal to each other will be zero.

From the eight equations independent of each other, it is able to deduce the four equilibrium equations and four continuity equations, including the fluid continuity equation, current continuity equation, and force equilibrium equation and so forth.

\begin{table}[h]
\centering
\caption{Eight equilibrium and continuity equations in the gravitational and electromagnetic fields.}
\begin{tabular}{@{}lll@{}}
\hline\noalign{\smallskip}
equation                                 &   definition                              &  subspace           \\
\noalign{\smallskip}\hline\noalign{\smallskip}
fluid continuity equation                &   $N_{10} = 0$                            &  $\mathbb{H}_g$     \\
torque continuity equation               &   $N_{10}^\textrm{i} = 0$                 &  $\mathbb{H}_g$     \\
force equilibrium equation               &   $\textbf{N}_1^\textrm{i} = 0$           &  $\mathbb{H}_g$     \\
precession equilibrium equation          &   $\textbf{N}_1 = 0$                      &  $\mathbb{H}_g$     \\
current continuity equation              &   $\textbf{N}_{20} = 0$                   &  $\mathbb{H}_{em}$  \\
second-torque continuity equation        &   $\textbf{N}_{20}^\textrm{i} = 0$        &  $\mathbb{H}_{em}$  \\
second-force equilibrium equation        &   $\textbf{N}_2^\textrm{i} = 0$           &  $\mathbb{H}_{em}$  \\
second-precession equilibrium equation   &   $\textbf{N}_2 = 0$                      &  $\mathbb{H}_{em}$  \\
\noalign{\smallskip}\hline
\end{tabular}
\end{table}

\section{Equilibrium or continuity equations}

In the octonion space $\mathbb{O}$ , when the octonion force is equal to zero, it is capable of deducing eight equilibrium and continuity equations, including the force equilibrium equation, fluid continuity equation, and current continuity equation in the classical field theory. These equations will be influenced by the gravitational strength, electromagnetic strength, torque, magnetic moment, and spatial dimension and others. These eight equations are derived from one single equation, $\mathbb{N} = 0$ , so that these eight equations are essentially the same (Table 3).

\subsection{Quaternion space}

In the quaternion space $\mathbb{H}_g$ for the gravitational fields, there are four equilibrium or continuity equations. Each of them connects to the angular momentum, including the derivative of angular momentum, divergence of angular momentum, and curl of angular momentum and so forth. Obviously, the equilibrium equations are relative to the derivative of angular momentum or the curl of angular momentum and other vectors. The continuity equations are relevant to the divergence of angular momentum and other scalars (Table 4).

1) Force. In the force equilibrium equation, $\textbf{N}_1^\textrm{i} = 0$, the physical quantities consist of the derivative of the curl $\textbf{W}_1$ of angular momentum, gradient of energy $W_{10}^\textrm{i}$ , curl of torque $\textbf{W}_1^\textrm{i}$ , gravitational strength, and electromagnetic strength and others. The equation can be further simplified into the force equilibrium equation in the classical mechanics. In the simplified equation derived from the force equilibrium equation, $\textbf{N}_1^\textrm{i} = 0$, the force comprises the inertial force, gravitational force, electromagnetic force, interaction force between the magnetic moment with magnetic field, interaction force between the electric moment with electric field, energy gradient, and external force and others. These force terms should meet the need of the force equilibrium equation simultaneously.

2) Power. In the fluid continuity equation, $N_{10} = 0$, the physical quantities consist of the divergence of the curl $\textbf{W}_1$ of angular momentum and the derivative of energy $W_{10}^\textrm{i}$ and others. The equation can be further reduced into the fluid continuity equation in the classical mechanics. In the simplified equation of the fluid continuity equation, $N_{10} = 0$, the physical quantities include the divergence of linear momentum and the derivative of mass and others. These physical quantities should simultaneously satisfy the requirement of the fluid continuity equation.

3) Torque derivative. In the precession equilibrium equation, $\textbf{N}_1 = 0$, the physical quantities cover the curl of the curl $\textbf{W}_1$ of angular momentum, derivative of torque $\textbf{W}_1^\textrm{i}$ , gradient of the divergence $W_{10}$ of angular momentum, gravitational strength, and electromagnetic strength and others. These physical quantities must satisfy the requirements of the precession equilibrium equation simultaneously. The precession equilibrium equation can be further reduced to a comparatively simple equation. The latter can derive some precessional angular velocities of particles in the gravitational and electromagnetic fields.

4) Torque divergence. In the torque continuity equation, $N_{10}^\textrm{i} = 0$, the physical quantities include the divergence of torque $\textbf{W}_1^\textrm{i}$ and the derivative of the divergence $W_{10}$ of angular momentum and others. The torque continuity equation can be further degenerated into a comparatively simple equation. The latter can infer the relationship between the external torque and vortices of the vortex street in the fluids.

In the quaternion space $\mathbb{H}_g$ for the gravitational fields, the four equilibrium and continuity equations are capable of revealing some physical properties of the gravitational substance.

\begin{table}[h]
\centering
\caption{Comparison of characteristics between the equilibrium equation and continuity equation.}
\begin{tabular}{@{}lllll@{}}
\hline\noalign{\smallskip}
item                          &   equilibrium equation                                &   continuity equation                                \\
\noalign{\smallskip}\hline\noalign{\smallskip}
equation                      &   vector equation                                     &   scalar equation                                    \\
operation                     &   derivative of a vector,                             &   derivative of a scalar,                            \\
                              &   curl of vectors                                     &   divergence of vectors                              \\
field strength                &   playing a major role in $\mathbb{H}_g$ ,            &   playing a secondary role                           \\
                              &   playing a secondary role in $\mathbb{H}_{em}$       &   in $\mathbb{H}_g$ and $\mathbb{H}_{em}$            \\
\noalign{\smallskip}\hline
\end{tabular}
\end{table}

\subsection{Second-quaternion space}

In the subspace $\mathbb{H}_{em}$ for the electromagnetic fields, there are two equilibrium equations, and two continuity equations. Each of them associates with the magnetic moment, including the derivative of magnetic moment, divergence of magnetic moment, and curl of magnetic moment and so on. Apparently, the equilibrium equations are relevant to the derivative of magnetic moment or the curl of magnetic moment and other vectors. The continuity equations are relative to the divergence of magnetic moment and other scalars.

1) Second-force. In the second-force equilibrium equation, $\textbf{N}_2^\textrm{i} = 0$, the physical quantities comprise the derivative of the curl $\textbf{W}_2$ of magnetic moment, gradient of the second-energy $\textbf{W}_{20}^\textrm{i}$ , curl of the second-torque $\textbf{W}_2^\textrm{i}$, gravitational strength, and electromagnetic strength and others. These physical quantities must satisfy the requirement of the second-precession equilibrium equation simultaneously. Further the equation can be simplified into one comparatively simple equation. In the simplified equation of the second-precession equilibrium equation, $\textbf{N}_2^\textrm{i} = 0$, the physical quantities consist of the derivative of electric current and the gradient of electric charge and so forth. And it can deduce one mechanism for the transport of droplets and the macroscopic surface charge gradients.

2) Second-power. In the current continuity equation, $\textbf{N}_{20} = 0$, the physical quantities consist of the divergence of the curl $\textbf{W}_2$ of magnetic moment and the derivative of second-energy $\textbf{W}_{20}^\textrm{i}$ and others. Further the equation can be degenerated into the current continuity equation in the classical mechanics. In the simplified equation derived from the current continuity equation, $\textbf{N}_{20} = 0$, the physical quantities include the divergence of electric current and the derivative of electric charge and others. These physical quantities should meet the need of the current continuity equation simultaneously.

3) Second-torque derivative. In the second-precession equilibrium equation, $\textbf{N}_2 = 0$, the physical quantities consist of the curl of the curl $\textbf{W}_2$ of magnetic moment, derivative of the second-torque $\textbf{W}_2^\textrm{i}$ , gradient of the divergence $\textbf{W}_{20}$ of magnetic moment, gravitational strength, and electromagnetic strength and others. These physical quantities will satisfy the requirement of the second-precession equilibrium equation simultaneously. Further the second-precession equilibrium equation can be simplified into one comparatively simple equation. The latter can be applied to explain the physical phenomena relevant to the direct and inverse piezoelectric effects.

4) Second-torque divergence. In the second-torque continuity equation, $\textbf{N}_{20}^\textrm{i} = 0$, the physical quantities comprise the divergence of second-torque $\textbf{W}_2^\textrm{i}$ and the derivative of the divergence $\textbf{W}_{20}$ of magnetic moment and others. Further the second-torque continuity equation can be reduced into a comparatively simple equation. The latter can reveal some relationships between the external second-torque and vortices of the electric-current, in the conductors and magneto-fluids.

In the subspace $\mathbb{H}_{em}$ for the electromagnetic fields, the four equilibrium or continuity equations are able to explore some physical properties of the electromagnetic substance.

As one of eight equilibrium and continuity equations, the equilibrium equation discussed in the following context can be written as, $\textbf{N}_2 = 0$, which is called as the second-precession equilibrium equation temporarily. From the second-precession equilibrium equation, it is able to reason out the direct and inverse piezoelectric effects and others relevant to the electromagnetic materials in the subspace $\mathbb{H}_{em}$.

In a word, there are many applications and examples relevant to the eight equilibrium and continuity equations, in the gravitational and electromagnetic fields (Table 5). The fluid continuity equation, force equilibrium equation, and current continuity equation were verified, and these three continuity or equilibrium equations were applied in many domains of science. Further, the precession equilibrium equation is used to study the physical property of Larmor precessional angular velocity (see Ref.[32]) and so forth. One can utilize the second-force equilibrium equation to research the transport of droplet (see Ref.[34]) and macroscopic surface charge gradient (see Ref.[35]). We may apply the second-precession equilibrium equation to investigate the direct piezoelectric effect (see Ref.[2]) and inverse piezoelectric effect (see Ref.[3]). Moreover, the torque continuity equation is able to explore the physical properties of vortex street and so on. The second-torque continuity equation is capable of researching the physical characteristics of magneto-fluids and others.

\begin{table}[t]
\centering
\caption{Some applications and examples relevant to the eight equilibrium and continuity equations in the gravitational and electromagnetic fields.}
\begin{tabular}{@{}lll@{}}
\hline\noalign{\smallskip}
equations                                &   applications                            &  examples           \\
\noalign{\smallskip}\hline\noalign{\smallskip}
fluid continuity equation                &   (verified)                              &  (many)             \\
torque continuity equation               &   vortex street                           &  ?                  \\
force equilibrium equation               &   (verified)                              &  (many)             \\
precession equilibrium equation          &   Larmor precession                       &  Ref.[32]           \\
current continuity equation              &   (verified)                              &  (many)             \\
second-torque continuity equation        &   magneto-fluid                           &  ?                  \\
second-force equilibrium equation        &   transport of droplet                    &  Ref.[34]           \\
                                         &   macroscopic surface charge gradient     &  Ref.[35]           \\
second-precession equilibrium equation   &   direct piezoelectric effect             &  Ref.[2]            \\
                                         &   inverse piezoelectric effect            &  Ref.[3]            \\
\noalign{\smallskip}\hline
\end{tabular}
\end{table}

\begin{table}[b]
\centering
\caption{The measurement experiments with respect to the theoretical analysis. In the first type of induced mode, when the derivative of electric moments is set up, the electric current can be induced and generated. This had been observed in the experiments of direct piezoelectric effect. When some dielectrics are deformed by the external forces in a certain direction, the polarization will occur inside the dielectrics. Meanwhile there are the positive and negative opposite charges on their two opposite surfaces. In the second type of induced mode, in case the electric current is preseted, the derivative of electric moments can be induced and produced within the dielectrics. This had been measured in the experiments of inverse piezoelectric effect.}
\begin{tabular}{@{}lll@{}}
\hline\noalign{\smallskip}
induced mode  &   experiment content                                                                       &    physical effect                  \\
\noalign{\smallskip}\hline\noalign{\smallskip}
first type    &   The derivative of electric moment, $\partial_0 ( k_{eg}^2 \textbf{L}_2^\textrm{i} )$,    &    direct piezoelectric effect      \\
              &   is able to induce and produce the electric current.                                      &                                     \\
second type   &   The presetting electric current, $\textbf{S}$, can induce                                &    inverse piezoelectric effect     \\
              &   and generate the derivative of electric moment.                                          &                                     \\
\noalign{\smallskip}\hline
\end{tabular}
\end{table}

\section{Second-precession equilibrium equation}

In the octonion space, if the octonion force is equal to zero, it is able to derive eight equilibrium or continuity equations, especially the second-precession equilibrium equation, $\textbf{N}_2 = 0$. Under the influences of the gravitational strength, electromagnetic strength, torque, and magnetic moment and others, it is capable of inferring the direct and inverse piezoelectric effects of the electromagnetic substance in the subspace $\mathbb{H}_{em}$ , from the second-precession equilibrium equation, $\textbf{N}_2 = 0$.

The second-precession equilibrium equation, $\textbf{N}_2 = 0$, can be expanded into (see Ref.[32]),
\begin{eqnarray}
0 = && ( \textbf{g} \circ \textbf{W}_{20} / v_0 + \textbf{g} \times \textbf{W}_2 / v_0 + \textbf{b} \circ \textbf{W}_{20}^\textrm{i} + \textbf{b} \times \textbf{W}_2^\textrm{i} ) / v_0
\nonumber
\\
&&
+ ( W_{10} \textbf{E} / v_0 + \textbf{E} \times \textbf{W}_1 / v_0 + W_{10}^\textrm{i} \textbf{B} + \textbf{B} \times \textbf{W}_1^\textrm{i} ) / v_0
\nonumber
\\
&&
+ (- \nabla \circ \textbf{W}_{20} - \nabla \times \textbf{W}_2 + \partial_0 \textbf{W}_2^\textrm{i} ) ~,
\end{eqnarray}
where $\textbf{W}_2 \approx v_0 k_p \textbf{P}$ . $k_p = k - 1$ . And $k$ is the spatial dimension of radius vector, $\textbf{r} = \Sigma r_k \textbf{d}_k$ , in the quaternion space $\mathbb{H}_g$ . In the octonion space, the above equation can be reduced to a few simple cases.

In case the contribution of field strength can be neglected, the above equation will be simplified into,
\begin{eqnarray}
- \nabla \circ \textbf{W}_{20} - \nabla \times \textbf{W}_2 + \partial_0 \textbf{W}_2^\textrm{i} = 0 ~.
\end{eqnarray}

The above means that the existence of the curl of electric current, $\nabla \times \textbf{W}_2$ , can exert an impact on the physical quantities, $\partial_0 \textbf{W}_2^\textrm{i}$ and $\nabla \circ \textbf{W}_{20}$ . The variations of the derivative of magnetic moment, curl of electric moment, and divergence of electric moment and others, within the two physical quantities, $\partial_0 \textbf{W}_2^\textrm{i}$ and $\nabla \circ \textbf{W}_{20}$ , are able to result in the fluctuations of the magnetic moment and electric moment, altering the electric charge, magnetism, and resistance and so forth. On the contrary, the two physical quantities, $\partial_0 \textbf{W}_2^\textrm{i}$ and $\nabla \circ \textbf{W}_{20}$ , have an influence on the curl of electric current, $\nabla \times \textbf{W}_2$ . The external force may make an effect on the fluctuations of magnetic moment and electric moment, inducing the curvilinear flow of electric current. That is, the varying physical quantities, $\textbf{W}_2^\textrm{i}$ (or $\textbf{W}_{20}$) and $\textbf{W}_2$ , are able to excite each other. This inference can be applied to explain some physical effects, including the piezoelectric effects, piezomagnetic effects, piezoresistive effects and so forth.

Further, when the contribution of the derivative of magnetic moment and divergence of electric moment can be neglected, the above equation will be degenerated into,
\begin{eqnarray}
\partial_0 ( - v_0 \nabla \times \textbf{L}_2^\textrm{i} ) - \nabla \times \textbf{W}_2 = 0 ~,
\end{eqnarray}
or
\begin{eqnarray}
\nabla \times ( \partial_0 \textbf{L}_2^\textrm{i} + k_p \textbf{P} ) = 0 ~.
\end{eqnarray}

By multiplying the coefficient of $k_{eg}^2$ , the above equation can be rewritten as,
\begin{eqnarray}
\partial_0 ( k_{eg}^2 \textbf{L}_2^\textrm{i} ) + k_p \textbf{S} = \textbf{K} ~,
\end{eqnarray}
where $k_{eg}^2 \textbf{P} = \textbf{S}$ , with $\textbf{S}$ being the electric current. $k_{eg}^2 \textbf{L}_2^\textrm{i}$ is the electric moment. $\textbf{K}$ is one vector, in the subspace $\mathbb{H}_{em}$ , to satisfy the requirement of $\nabla \times \textbf{K} = 0$ . And $\textbf{K}$ may be equal to zero sometimes.

According to the above equation, there are two types of induce modes within the electromagnetic substance (Table 6). In the first type of induce mode (Table 7), the derivative of electric moment, $\partial_0 ( k_{eg}^2 \textbf{L}_2^\textrm{i} )$ , may excite the electric current, $\textbf{S}$ . In the second type of induce mode (Table 8), the electric current, $\textbf{S}$ , will induce the derivative of electric moment, $\partial_0 ( k_{eg}^2 \textbf{L}_2^\textrm{i} )$ .

The above inference can be considered as the underlying mechanism of direct and inverse piezoelectric effects. It should be noted that the spatial dimension of radius vector has a direct influence on the piezoelectric effects. According to Eq.(16), there are the two- or three-dimensional piezoelectric effects. However, there exists $k_p = 0$, in case $k = 1$. It implies that there is no one-dimensional direct or inverse piezoelectric effect, in the strict sense. When the piezoelectric nano-wires (or piezo-proteins) are folded into two- or three-dimensional shapes, it must be able to achieve higher piezoelectric coefficients than that of one-dimensional shape.

Apparently, the inferences derived from Eq.(12) should be much more complicated than that derived from Eq.(16). It means that there may be a few more complicated physical phenomena relevant to piezoelectric effects than ever before. And it is beneficial to further understand the variation law of electric currents and derivative of electric moments, in terms of some quartz crystals and others.

These inferences can be applied to the bio-mechanics, explaining some biological characteristics. a) According to the second-procession equilibrium equation, the electric current and derivative of electric moments can excite each other. On a relatively stable biological structure, the derivative of electric moments is able to induce the electric current, constituting an ion channel. In terms of the biological mechanisms capable of relative motions, the fluids and other media can form ion channels directly. It has been confirmed in the experiments of piezo-proteins (see Refs.[30] and [31]). b) According to the second-force equilibrium equation, the charge gradient and derivative of electric current can excite each other. Near the interfaces of fluids, the charge gradients of the ion channels can produce the derivatives of electric currents (see Refs.[34] and [35]), and vice versa. This can be used to explain the physical phenomena in the experiments related to bioelectric currents, such as the neural current, muscle current, skin current, cardiac current and so forth. Due to the existence of deformation (or derivative of electric moment), ion channel (or charge gradient) and pulse current (or derivative of electric current), these biological experiments involve the second-force equilibrium equation and second-procession equilibrium equation simultaneously.

\begin{table}[t]
\centering
\caption{The comparison between the theoretical prediction in Eq.(16) and experimental results of the piezoelectric effects. The derivative of electric moment, $\partial_0 ( k_{eg}^2 \textbf{L}_2^\textrm{i} )$ , and electric current, $\textbf{S}$ , must meet the need of Eq.(16), according to the electromagnetic theory. In the first type of induced mode, most of theoretical predictions, derived from Eq.(16), agree with the experimental results of the direct piezoelectric effect. However a few items of relevant experiments are still lacking at present. The subsequent experiments will continue to examine these theoretical predictions.}
\begin{tabular}{@{}lll@{}}
\hline\noalign{\smallskip}
physical quantity                                                                  &   theoretical prediction                                &   experiment result  \\
\noalign{\smallskip}\hline\noalign{\smallskip}
derivative of electric moment, $\partial_0 ( k_{eg}^2 \textbf{L}_2^\textrm{i} )$   &   presetting                                            &   presetting         \\
electric current                                                                   &   induced                                               &   obeying            \\
orientation of electric current                                                    &   $- \partial_0 ( k_{eg}^2 \textbf{L}_2^\textrm{i} )$   &   obeying            \\
amplitude of electric current                                                      &   meet the need of Eq.(16)                              &     ?                \\
rising/falling edge of charge curve                                                &   triggered by the derivative of electric moments       &   obeying            \\
increment of electric current                                                      &   variation of the derivative of electric moments       &     ?                \\
\noalign{\smallskip}\hline
\end{tabular}
\end{table}

\begin{table}[b]
\centering
\caption{Comparison of characteristics between the theoretical predictions in Eq.(16) with experiment results of the inverse piezoelectric effect. The electric current, $\textbf{S}$ , and derivative of electric moment, $\partial_0 ( k_{eg}^2 \textbf{L}_2^\textrm{i} )$ , must satisfy the requirement of Eq.(16), according to the electromagnetic theory. In the second type of induced mode, most of theoretical predictions, derived from Eq.(16), are in accord with the experimental results of the inverse piezoelectric effect. However several contents of relevant experiments are still missing up to now. The future experiments will further improve the detection accuracy, examining these theoretical predictions.}
\begin{tabular}{@{}lll@{}}
\hline\noalign{\smallskip}
physical quantity                                           &   theoretical prediction                     &   experiment result                \\
\noalign{\smallskip}\hline\noalign{\smallskip}
electric current, $\textbf{S}$                              &   presetting                                 &   presetting                       \\
derivative of electric moments                              &   induced                                    &   obeying                          \\
orientation of the derivative of electric moments           &   $- \textbf{S}$                             &   obeying                          \\
amplitude of the derivative of electric moments             &   satisfy the requirement of Eq.(16)         &     ?                              \\
rising/falling edge of electric moment curve                &   triggered by the electric current          &   obeying                          \\
increment of the derivative of electric moments             &   variation of the electric current          &     ?                              \\
\noalign{\smallskip}\hline
\end{tabular}
\end{table}

\section{Experiment proposal}

According to Eq.(16), there are two types of induce modes, in the electromagnetic substance. In the first type of induce mode, the derivative of electric moments will generate the electric current. It can be applied to explain the physical phenomena of direct piezoelectric effect. In the second type of induce mode, the existence of electric current will produce the derivative of electric moments. And it is able to describe the physical characteristics of inverse piezoelectric effect. Apparently, the electric current and varying electric moment can excite each other.

Most theoretical inferences, derived from Eq.(16), are consistent with the physical characteristics of the direct and/or inverse piezoelectric effects. However, a few theoretical inferences of Eq.(16) have not been confirmed by experiments. We can moderately improve the experimental scheme of the direct and/or inverse piezoelectric effects, verifying several theoretical inferences of Eq.(16) that have not been tested.

1) Direct piezoelectric effect. More than 20 types of crystals in nature are possessed of piezoelectric effects. The quartz crystal is the most representative and widely used. The quartz crystal is anisotropic. By cutting along different directions, it is able to obtain different geometric-cut quartz wafers. Apparently, the piezoelectric constants of quartz crystals are not equal in all directions.

In the Table 7, the derivative of electric moment, $\partial_0 ( k_{eg}^2 \textbf{L}_2^\textrm{i} )$, is one independent variable, while the electric current, $\textbf{S}$ , is one dependent variable. According to Eq.(16),  when the derivative of electric moment, $\partial_0 ( k_{eg}^2 \textbf{L}_2^\textrm{i} )$, possesses one single frequency and $\textbf{K} = 0$, the frequency of electric current, $\textbf{S}$ , will be the same as that of electric moment, $\partial_0 ( k_{eg}^2 \textbf{L}_2^\textrm{i} )$. In particular, if the derivative of electric moment, $\partial_0 ( k_{eg}^2 \textbf{L}_2^\textrm{i} )$, is one sinusoidal function, the electric current, $\textbf{S}$ , must be one cosine function.

When the quartz crystal is not subjected to the external force, the positive charge center of quartz crystal coincides with its negative charge center. As a result, the total electric moment of the whole quartz crystal is equal to zero, and the crystal surfaces are not charged. When the quartz crystal is subjected to the external force, the variation of crystal length leads to the fluctuation of electric moments. Due to the deformation of quartz crystal, the center of positive charges no longer coincides with that of negative charges.

The quartz crystal is capable of producing one tiny electric current, when it is subjected to external force. The variation of electric current leads to the generation of electric charge. Consequently, there are different electric charges on the surfaces of both ends of the crystal. By improving the experimental scheme related to the direct piezoelectric effect of quartz crystal, we can more accurately measure the corresponding relationship between the derivative curve of electric moment and the curve of induced electric current.

2) Inverse piezoelectric effect. When the quartz crystal is not affected by the external electric current, the positive charge center of quartz crystal coincides with the negative charge center. The total electric moment of the whole quartz crystal is zero, and the crystal surface is not charged. When the quartz crystal is subjected to one external electric current, the electric moment will change. The variation of electric moment leads to the mutation of crystal length.

In the Table 8, the electric current, $\textbf{S}$ , is one independent variable, while the derivative of electric moment, $\partial_0 ( k_{eg}^2 \textbf{L}_2^\textrm{i} )$, is one dependent variable. According to Eq.(16), if the electric current, $\textbf{S}$ , possesses one single frequency and $\textbf{K} = 0$, the frequency of the derivative of electric moment, $\partial_0 ( k_{eg}^2 \textbf{L}_2^\textrm{i} )$, must be identical to that of electric current, $\textbf{S}$ . In particular, if the electric current, $\textbf{S}$ , is one sinusoidal function, the derivative of electric moment, $\partial_0 ( k_{eg}^2 \textbf{L}_2^\textrm{i} )$, will be one cosine function.

When the quartz crystal is subjected to one external electric current, it is able to produce a tiny derivative of electric moment. The variation of the derivative of electric moments is capable of altering the electric moment. Consequently, the variation of electric moments leads to the fluctuation of crystal length. By improving the experimental scheme related to the inverse piezoelectric effect of quartz crystal, we can more accurately measure the corresponding relationship between the electric current curve and the derivative curve of electric moments.

Further, by means of the direct and inverse piezoelectric effects, it is able to produce the high-precision oscillation frequencies. The quantity of induced electric charges varies with the external force. The induced vibration frequency of the quartz crystal varies with the external voltage. Making use of the direct and inverse piezoelectric effects, we can study the corresponding relationship between the electric current curve and the derivative curve of electric moments.

The verification of experiment proposal will be helpful to further understand the physical properties of the electric current and the derivative of electric moments.

\section{Conclusions and discussions}

The octonions are capable of describing the physical properties of electromagnetic and gravitational fields simultaneously, including the octonion field strength, field source, linear momentum, angular momentum, torque, force and so forth. The octonion space $\mathbb{O}$ can be separated into a few subspaces independent of each other, in particular two subspaces, $\mathbb{H}_g$ and $\mathbb{H}_{em}$ . The subspace $\mathbb{H}_g$ is one quaternion space, which is fit to research the physical properties of gravitational fields. Meanwhile, the second subspace $\mathbb{H}_{em}$ is appropriate to explore the physical properties of electromagnetic fields. Under some conditions, the second subspace $\mathbb{H}_{em}$ can be transformed into the second quaternion space $\mathbb{H}_e$ . Apparently, the second quaternion space $\mathbb{H}_e$ is independent of the quaternion space $\mathbb{H}_g$ .

When the octonion force is equal to zero, it is able to achieve eight equilibrium and continuity equations independent of each other, from $\mathbb{N} = 0$. Four of these eight equations are situated in the quaternion space $\mathbb{H}_g$ , including the force equilibrium equation, fluid continuity equation, precession equilibrium equation, and torque continuity equation. The rest of these eight equations are located in the second subspace $\mathbb{H}_{em}$ , including the second-force equilibrium equation, current continuity equation, second-precession equilibrium equation, and second-torque continuity equation. Furthermore, when the contribution of field strength is ignored, we can obtain the relationship between the derivative of electric moments and the electric current, from the second-precession equilibrium equation. This inference can be considered as the underlying mechanism of the direct and inverse piezoelectric effects.

According to Eq.(16), the external force is able to alter the length of quartz crystals, compelling the electric moment of quartz crystals to twist to a certain extent. The torsion of the electric moment of quartz crystals can vary the derivative of electric moments. Furthermore the existence of the derivative of electric moments will induce the generation of weak electric current, producing different electrical charges on the surfaces of quartz crystals. This inference can be considered as the underlying mechanism of direct piezoelectric effects, in terms of the quartz crystals and others.

On the other hand, according to Eq.(16), the external electric currents are capable of altering the derivative of electric moments of the quartz crystals. The existence of the derivative of electric moments leads to the torsion of the electric moments of quartz crystals. The torsion of the electric moments of quartz crystals results in the variation of the length of quartz crystals. This inference can be considered as the underlying mechanism of inverse piezoelectric effects, in terms of the quartz crystals and others. Apparently, the direct and inverse piezoelectric effects both obey one single formula, Eq.(16).

It is noteworthy that this paper merely discusses some simple cases associated with the interplay between the second-torque $\textbf{W}_2^\textrm{i}$ and the curl of magnetic moment, $\textbf{W}_2$. However, it has clearly shown that the second-torque can have a direct impact on the curl of magnetic moment, including the amplitudes and orientations and so forth. In particular, the derivative of electric moments and the electric current can excite each other. And it can be applied to explain the physical phenomena relevant to the direct and inverse piezoelectric effects. In the following study, it is going to further explore the influences of not only the second-torque, $\textbf{W}_2^\textrm{i}$, but also the divergence of magnetic moment, $\textbf{W}_{20}$, on the curl of magnetic moment, $\textbf{W}_2$, theoretically, by means of the second-procession equilibrium equation, Eq.(12). In the experiments, we attempt to measure the contribution of some terms of the divergence of magnetic moment, $\textbf{W}_{20}$, on the curl of magnetic moment, $\textbf{W}_2$. On the basis of Eq.(16), it is possible to further research the contribution of other terms to the direct and inverse piezoelectric effects.

\section*{Acknowledgement}
The author is indebted to the anonymous referees for their valuable comments on the previous manuscripts. This project was supported partially by the National Natural Science Foundation of China under grant number 60677039.

\end{document}